\documentclass[12pt,preprint]{aastex}

\usepackage{natbib}
\usepackage{amsfonts}

\usepackage{color}
\usepackage{ulem}

\shorttitle{Coronal Limb EUV Waves and Particles}
\shortauthors{Kozarev K. A., Korreck K. E., Lobzin V.V., Weber M. A., Schwadron N. A.}

\begin{document}


\title{Off-limb Solar Coronal Wavefronts From SDO/AIA EUV Observations---Implications For Particle Production}

%

\author{K. A. Kozarev\altaffilmark{1}}
\affil{Astronomy Department, Boston University,
    Boston, MA 02215}
\email{kamen@bu.edu}

\author{K. E. Korreck}
\affil{Harvard-Smithsonian Center for Astrophysics,
    Cambridge, MA 02138}

\author{V. V. Lobzin}
\affil{School of Physics, University of Sydney, 
 NSW 2006, Australia}

\author{M. A. Weber}
\affil{Harvard-Smithsonian Center for Astrophysics,
 Cambridge, MA 02138}
\and

\author{N. A. Schwadron}
\affil{Institute for the Study of Earth, Oceans, and Space, University of New Hampshire, 
 Durham, NH 03824}

\altaffiltext{1}{also at Harvard-Smithsonian Center for Astrophysics,
Cambridge, MA 02138}

\begin{abstract}
We derive kinematic properties for two recent solar coronal transient waves observed off the western solar limb with the Atmospheric Imaging Assembly (AIA) onboard the Solar Dynamics Observatory (SDO) mission. The two waves occurred over $\sim10$-min intervals on consecutive days---June 12 and 13, 2010. For the first time, off-limb waves are imaged with a high 12-s cadence, making possible detailed analysis of these transients in the low corona between $\sim1.1$--2.0 solar radii ($R_{s}$). We use observations in the 193 and 211~{\AA} AIA channels to constrain the kinematics of both waves. We obtain initial velocities for the two fronts of $\sim1287$ and $\sim736$ km~s$^{-1}$, and accelerations of $-1170$ and $-800$ m~s$^{-2}$, respectively. Additionally, differential emission measure analysis shows the June 13 wave is consistent with a weak shock. EUV wave positions are correlated with positions from simultaneous type II radio burst observations. We find good temporal and height association between the two, suggesting that the waves may be the EUV signatures of coronal shocks. Furthermore, the events are associated with significant increases in proton fluxes at 1~AU, possibly related to how waves propagate through the coronal magnetic field. Characterizing these coronal transients will be key to connecting their properties with energetic particle production close to the Sun.\\\\
\end{abstract}

\section{INTRODUCTION}

A growing body of theoretical and observational research suggests that charged solar energetic particles (SEPs) gain most of their energy at traveling shocks relatively close to the Sun \citep{Zank:2008}. Interplanetary shocks have been well studied with in-situ measurements near Earth and throughout the solar system \citep{Stone:1985,Forbes:2006}. Many SEP bursts observed close to Earth are not directly associated with Earth-detected shocks. This suggests that SEPs are accelerated much closer to the solar corona, possibly by shocks near the Sun \citep{Reiner:2007}. Coronal shocks could accelerate particles to very high energies in short periods \citep{Roussev:2004}. However, field and shock geometry are key parameters in the ability of shocks to accelerate particles regardless of the shock strength, especially near the Sun \citep{Giacalone:2006}.


Coronal shocks have been observed earlier\citep{Pick:2006,Nindos:2008}. \citet{Maia:2000} reported on fast coronal transients propagating with similar speeds in both radio and white light. \citet{Vourlidas:2003} used coronagraph observations to study a white-light coronal shock beyond 2.5 $R_S$. Recent results \citep[e.g.,][]{Gallagher:2010, Patsourakos:2009a, Patsourakos:2009b} have used EUV observations to show the intimate connection between EUV waves and CMEs. However, there is still considerable debate about how shocks appear in these observations. Additionally, a widely used means of characterizing coronal shock kinematics is observations of drifting metric radio emissions from the Sun (approximately 18--180 MHz). These type II radio bursts are associated with coronal shocks accelerating electrons that excite plasma radio emissions \citep{McLean:1985,Reiner:2003,Mancuso:2003}.

Ultra-high cadence EUV imaging observations of off-limb coronal waves are presented, for two recent solar eruptions on June 12 and June 13, 2010. We use the Atmospheric Imaging Assembly (AIA) instrument \citep{Title:2006} aboard the Solar Dynamics Observatory. The temporal resolution, $\sim$12~s, allows for following the evolution of impulsive features in the low corona (1.2--2.0~$R_S$)---a capability newly available in EUV imaging of the Sun. About 3 hours after the June 12 event and 2 hours after the June 13 event, elevated proton fluxes (~6.5 MeV) were observed at 1 AU, leading us to investigate the connection between the remote wave observations and in-situ particle fluxes. We combine simultaneous EUV wave and type II radio burst observations with a coronal magnetic field model to investigate the morphology, kinematics, thermal and density properties of the wavefronts, and their energetic particle production capability.


The Letter is structured as follows: In Section~2, we detail the AIA and radio observations used. In Section~3 the kinematics, morphology, and physical properties of the EUV transients are described. We summarize our findings in Section~4.

\section{OBSERVATIONS AND DATA ANALYSIS}

\subsection{EUV observations}

We used observations from two AIA channels peaking at 193(FeXII) and 211~{\AA}(FeXIV). We refer to them as the 193 and 211 channels throughout the paper. Both channels have $\sim12$~s cadence, $~6$-s lag between the two. The data were processed to level 1.5 using a standard AIA pipeline. Base difference images were produced from an average of ten subframes immediately preceding the events. Event movies can be found as online supplemental materials to this Letter. For temperature and density analysis of the June 13 event, we use six EUV AIA channels dominated by Fe lines (details are given below).

The first event occurred on June 12, 2010 above active region (AR) 11081 located close to the northwestern limb (N23W43). The EUV transient coincided with an M2.0 X-ray flare between 00:30--01:02 UT, peaking at 00:57 UT. We considered observations between 00:56--01:03~UT, the times during which we could detect and measure the coronal transient feature in the FOV of the AIA instrument. During this period a faint, but discernible front was launched roughly radially above the AR.

The second event occurred on June 13, 2010 above AR11079, on the southwestern limb (S25W84). It coincided with an M1.0 flare between 05:30-05:44~UT, peaking at 05:39 UT. In EUV an eruption started at 05:34~UT on the limb, turning into a CME loop propagating radially outward. At 05:37~UT, a hemispheric wavefront appeared (in both 193 and 211~channels) in front of the CME and separated from it, traveling in the same direction but markedly faster. The wave reached the AIA FOV edge at 05:42 UT, followed by the CME at 05:44 UT. 

\begin{figure}
\epsscale{0.8}
\plotone{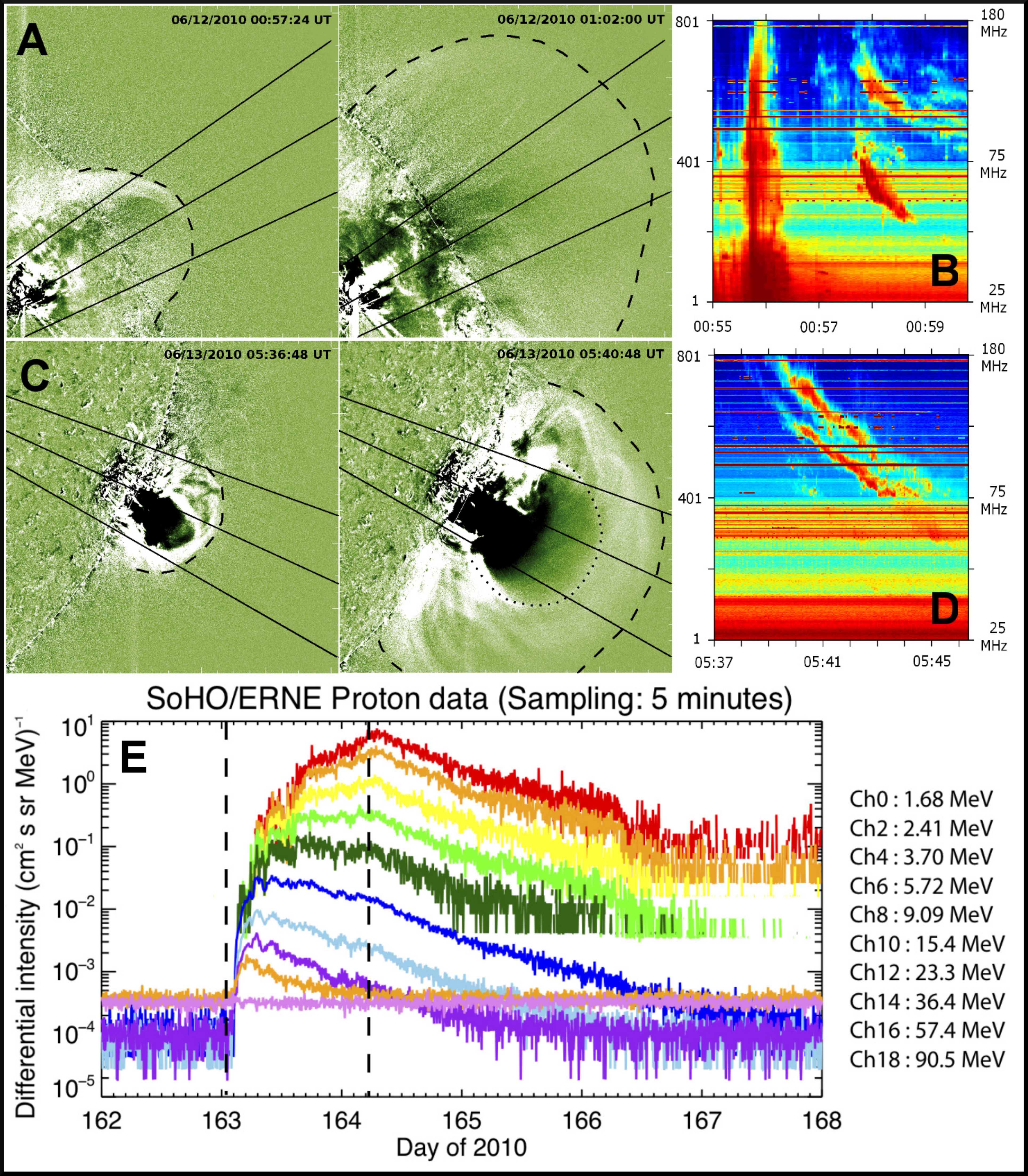}
\caption{Panels A and C: AIA/211 base difference images showing two stages of the June 12 and 13, 2010 coronal waves, respectively. The two frames in each panel are $\sim4$ minutes apart. Approximate positions of the wavefronts are in dashed black lines. Dotted lines in panel C outline the June 13 CME. The radial profiles along which velocity measurements were made are also shown. Panels B and D: radio spectra from Learmonth observatory for June 12 and 13, respectively. Panel B also shows a strong type III burst around 00:56~UT on June 12. Panel E: proton fluxes observed between June 11(DOY 162) and June 17(DOY 168), 2010 by the SOHO/ERNE instrument. Proton energies vary between 1.68-90.5 MeV. Vertical dashed lines show AIA waves onsets.\label{fig1}}
\end{figure}

In Figure \ref{fig1}, panels A and C show two base difference images in the 211 AIA channel of the June 12 and 13 events, respectively. Dashed lines trace the wavefronts. Although both events were only clearly visible in difference images, the second event was notably brighter, exhibiting lower velocities, as we show below. 

Since wave signatures were very faint, we made manual measurements. For each subframe in each event, the expanding wavefront edge was selected along three radial profiles close to the wave's nose. To reduce measurement errors, they were repeated ten times for each image sequence in both channels. We fitted second-order polynomials to measured positions in order to obtain front velocities and accelerations, using MPFIT routines \citep{Markwardt:2009} combined with a statistical bootstrapping technique \citep{Efron:1979}. Since the waves were very dim, we only managed meaningful observations for two profiles in each event. Third-order fits were also attempted, but did not produce significantly different results.

We also corrected for plane-of-sky projection of the wavefronts, assuming spherical waves propagating radially away from the Sun. Then, the brightest EUV emission is detected at front edges. We deprojected front positions by assuming $r=r'/\sin(\phi)$, where $r$ and $r'$ are the true and projected radial distances from the flare site, respectively, and $\phi$ is the AR heliographic longitude. Velocities and accelerations are presented in Table \ref{table1}.

\subsection{RADIO OBSERVATIONS}
Metric radio spectra were provided by the Learmonth Solar Radio Observatory (Western Australia). Type II bursts indicate electron acceleration by coronal shocks, which  may also accelerate protons and heavier ions. The Newkirk coronal electron density model \citep{Newkirk:1961} was used to relate the observed emissions to the height of the emission source.

Figure \ref{fig1}, panels B and D, show type II burst dynamic spectra. Multiple bands are visible for the June 12 type II radio burst, indicating that this event is rather complex. Both fundamental and harmonic emissions were observed for that event, starting at 00:57:45 UT. The harmonic emission persisted longer, until about 01:07 UT, but was too faint to be measured. A strong type III burst was also observed at 00:53 UT---an indication of an impulsive release of electrons in the corona. We separated two emission lanes in each spectrogram and fitted the peak emission frequencies.

We performed the same analysis for the June 13 radio burst, which started at 05:38:13 UT. The fundamental emission was barely discernible in the radio spectrogram (panel D). However, there were two parallel bands of harmonic emission.

\section{RESULTS AND DISCUSSION}

Figure \ref{fig1}, panel E, shows particle flux enhancements possibly associated with the coronal shocks observed by AIA. The in-situ particle measurements were made by the Energetic and Relativistic Nuclei and Electron \citep[ERNE;][]{Torsti:1995} instrument on SOHO. The time series of energetic protons (between 1.68--90.5 MeV) show an impulsive flux increase at all energies on June 12, followed by an additional increase in the low energies on June 13. Vertical dashed lines denote onsets of the EUV waves. Below we investigate EUV wave observations and radio shock properties in an attempt to characterize the solar sources of the elevated particle fluxes.

\subsection{Kinematics of the EUV waves and radio shocks}
Table \ref{table1} presents measured EUV wave front and radio shock kinematics. As described previously, we performed measurements of front edge positions along three linear profiles starting from the flare region (hereafter trials). These are denoted in roman numerals in the second column of the Table, together with the AIA channel. The third and fourth columns show initial velocities and acceleration, respectively, derived from second-order polynomial fits to the de-projected position measurements in two trials for each event. The rows in bold show trial averages for each wavelength, for each event.

For June 12, we obtained velocities of $\sim1275\pm44$~km~s$^{-1}$ for AIA/193 and $\sim1300\pm44$~km~s$^{-1}$ for AIA/211 channel. For June 13, we get $\sim731\pm22$~km~s$^{-1}$ for AIA/193 and $\sim741\pm31$~km~s$^{-1}$  for AIA/211 channel. The fits imply average decelerations of $-1000$~m~s$^{-2}$ for June 12 and $-800$~m~s$^{-2}$ for June 13.

\citet{Patsourakos:2010} studied the June 13 CME in EUV with AIA, between 05:34--05:43 UT. They fit circles to the expanding CME bubble and determined its kinematics. They found that the bubble front in the direction of propagation away from the solar limb accelerated to a maximum speed of ~400 km/s, after which it decelerated. They did not comment on the wave kinematics in that work.

\citet{Veronig:2010} studied a very similar dome-like CME and wave event off the eastern solar limb (seen from the STEREO-B spacecraft) with EUV observations They found upward expansion speeds of the dome-like wave of $\sim650$ km~s$^{-1}$. They also found that the EUV wavefront coincided with the white light transient observed by STEREO-B coronagraphs. This implies that the front edge of the white light emission may be caused by compressed elecron plasma behind the traveling shock. Quadrature position modeling was done by \citet{Patsourakos:2009b} to first show this connection.

\begin{figure}
\epsscale{0.8}
\plotone{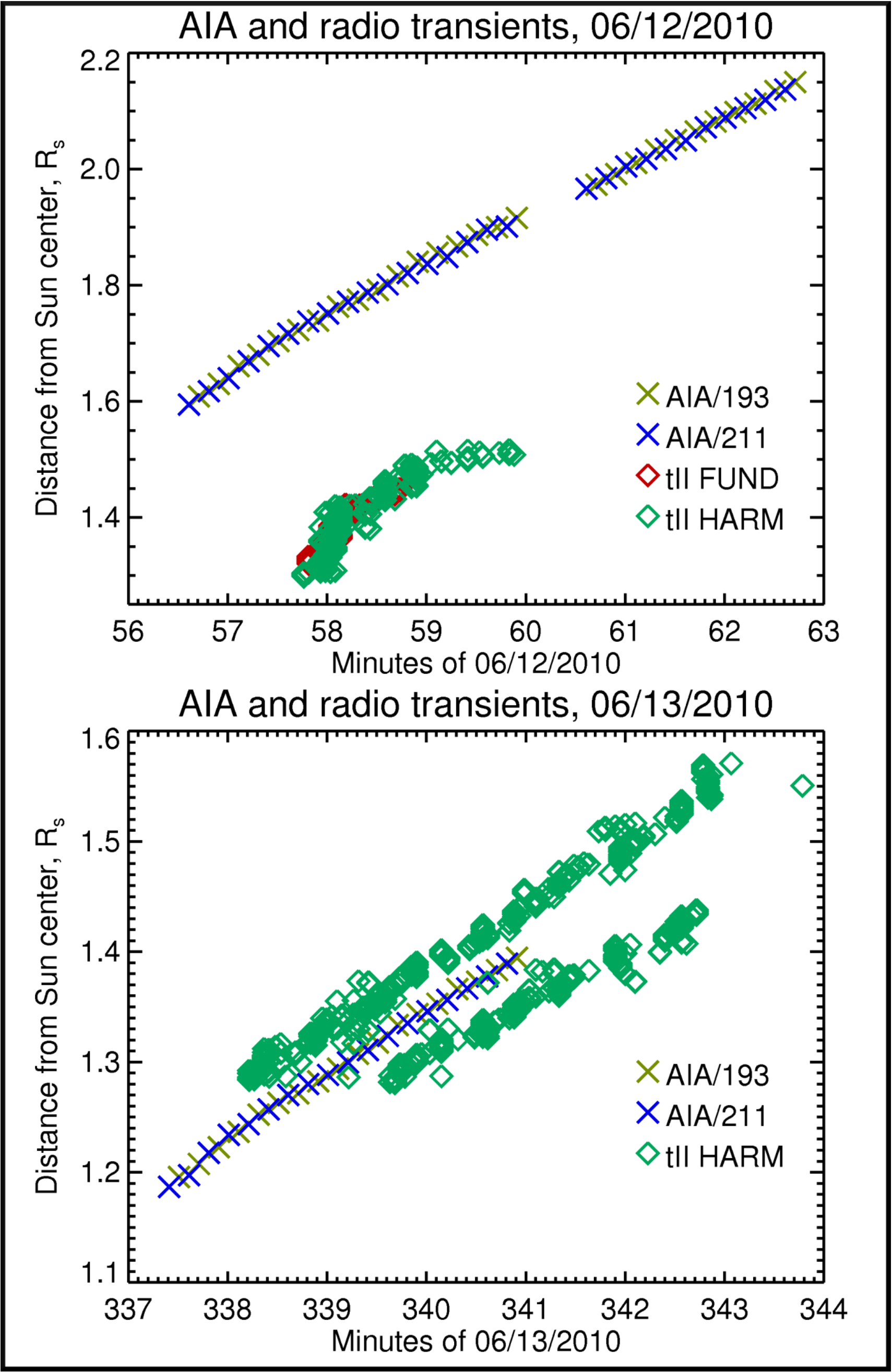}
\caption{Time-height profiles of June 12 (top) and 13 (bottom) EUV waves and radio shocks.  Wavefront positions from the lowest uncertainty trial (see Table \ref{table1}) from AIA/193 and 211 channels are shown as X-symbols. Diamonds denote shock positions estimated from radio type II burst observations with the Newkirk density model.\label{fig2}}
\end{figure}

Figure \ref{fig2} compares measured AIA EUV wavefront positions and estimated shock locations from radio observations. Wavefront positions measured in the 193 and 211 channels for the trial with lowest uncertainties are plotted as X-symbols. Diamonds denote radio shock positions. The June 12 radio emission occurred at lower heights than the EUV wavefront, suggesting electron acceleration away from the shock nose. Alternatively, the electron density model used might not apply for this case of open magnetic geometry (see Section~3.4). Radio emission started faster than the EUV wave, but decelerated completely by 01:00~UT, while the EUV wave continued to rise. By contrast, the radio emission on June 13 was split into two harmonic bands, which correlate very well with the EUV wave positions. This might imply local electron acceleration in front of and behind the nose of the traveling shock. Section~3.4 considers the coronal magnetic geometry in interpreting these observations.



\subsection{Temperature and density behavior of the EUV waves}

\begin{figure}
\epsscale{1.0}
\plotone{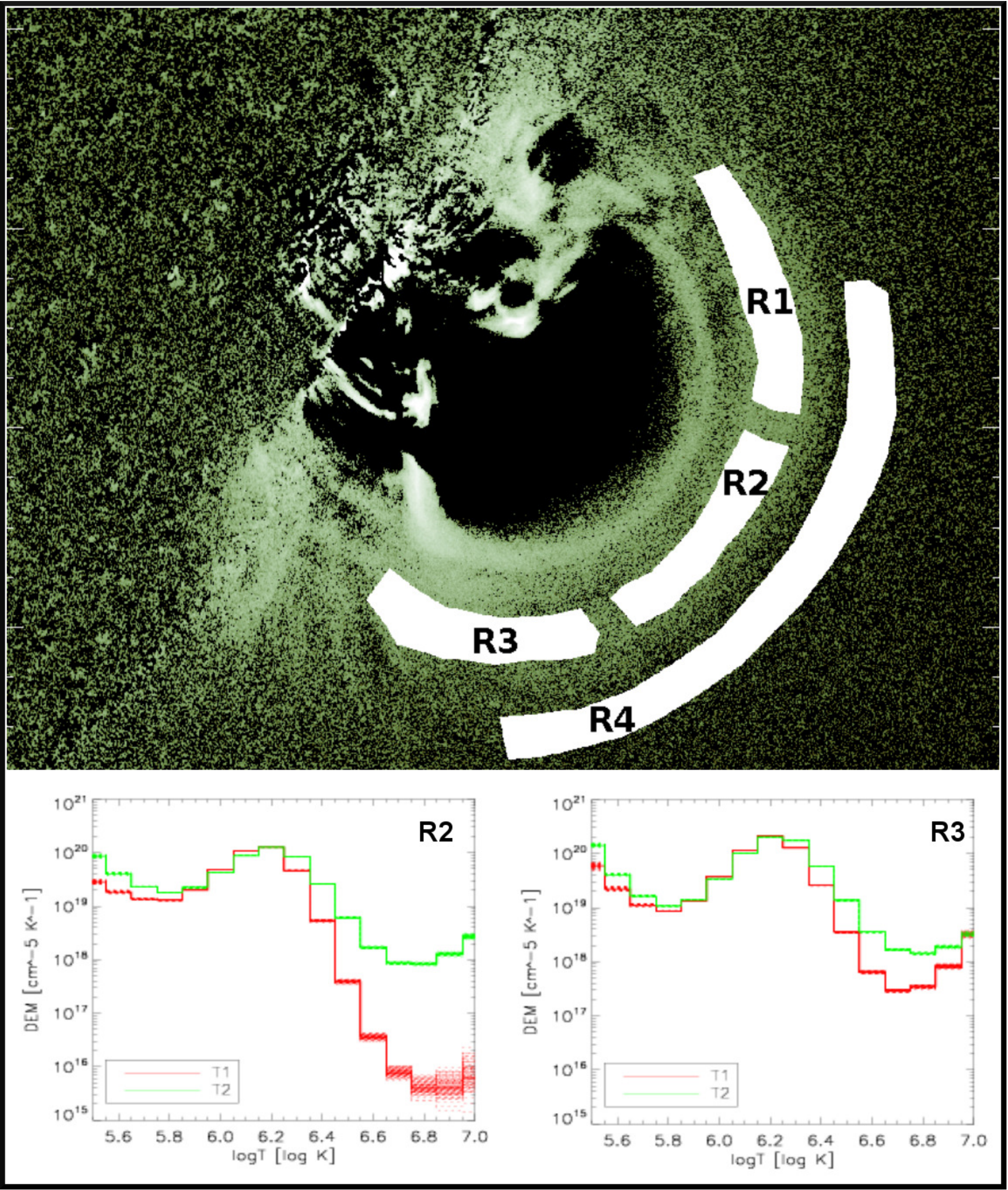}
\caption{Top - a snapshot of the June~13 event (base difference) with overlaid regions for which DEM solutions were attempted. Bottom - the DEM solutions for regions 2(left) and 3(right) as overlaid dotted histograms for the two times T1(red) and T2(green).\label{fig3}}
\end{figure}

To investigate the temporal and density properties of the EUV waves, we performed differential emission measure (DEM) analysis on the June 13 wave (we were not able to do so for the June 12 wave due to data constraints), using region-averaged pixel values in the six EUV Fe-dominated channels (94, 131, 171, 193, 211, 335~\AA). We hand-selected four regions (labeled R1-4 in top panel of Fig.\ref{fig3}) in two frames - 05:37:00~UT and 05:39:00~UT (hereafter T1 and T2) - corresponding to times before and during the wavefront passage. The first three regions were chosen to sample different parts of the wave; the fourth was chosen upstream of the wave for comparison. Calculations were done for 16 temperature bins between $logT=5.5-7.0$, following the Monte Carlo method as implemented in \citet{Weber:2004}. Results for regions 1 and 4 were not statistically significant, so our analysis was limited to times T1 and T2 in R2 and R3.

For each region (of approximately 10,000 pixels), time, and channel, we constructed mean intensity data and errors.  The mean observation sets were then solved for their DEM distributions. The DEM solutions we quote provide model intensities with the smallest $\chi^2$ fit to the data, when folded through the AIA responses. (See bottom panels of Fig.~\ref{fig3}.) We considered the relative degree of model fits versus the difference between T1 and T2 data.)

The DEMs for regions 2 and 3 are plotted in the bottom panels of Fig.\ref{fig3}, where T1 is shown in red and T2 is shown in green. The Monte Carlo analysis produces multiple solutions by varying the data by the errors, and these are represented as clouds of colored dotted lines with a very small spread. It can be seen that observations for T1 and T2 are significantly different. We find that the DEM temperature profile does not change appreciably from T1 to T2, for either region, but the overall emission measure increases.

To roughly estimate the jump in density, we consider a simple model. Assume that all measured intensity is emitted along the region's line-of-sight only from the wave-affected volume, i.e., no foreground nor background emission. Also assume no change in temperature. Then, since the integrated DEM is the full emission measure (EM) of the volume, we may estimate the density ratio as:

\begin{equation}
\frac{n_{e2}}{n_{e1}} \sim \frac{\sqrt{EM_2}}{\sqrt{EM_1}} \sim 
  \frac{\sqrt{\int \mathrm{DEM}_2(T) \mathrm{d}T}}{
        \sqrt{\int \mathrm{DEM}_1(T) \mathrm{d}T}}
\end{equation}

For region 2, we find that $n_{e2}/n_{e1} \simeq 1.18$, and for region 3, we find that $n_{e2}/n_{e1} \simeq 1.12$, consistent with weak coronal shocks. For a more sophisticated model that accounts for foreground and background emission, the density changes within the wave-affected volume would have to be {\it even larger} in order to generate the observed change in intensities.  Therefore, we find that $n_{e2}/n_{e1} \simeq 1.12$ is a {\it lower} limit.

\subsection{EUV wave morphology}
%

In both 211 and 193 channels, brightness increases downstream of the EUV wavefronts, relative to upstream. In the 211 channel this brightening is more pronounced, and is also more uniform throughout the downstream region. In the 193 channel, by contrast, the downstream material emits only close to the leading front. In both cases, ripples of emission behind the wavefronts expand as the fronts sweep through regions of upstream coronal plasma. These features persist until the transients leave the AIA FOV for both events. However, the downstream sheath emission on June 12 was dimmer than the emission on June 13 (where a CME bubble was seen).

\subsubsection{Lateral Overexpansion of the June 13 CME}

\citet{Patsourakos:2010} studied the CME of June 13, and found a strong lateral overexpansion of the CME bubble in the first ~4 minutes, after which the bubble expansion became equal in the radial and lateral directions (see top panel in their Fig.4.) Recently reported 3D numerical MHD simulations of coronal CME propagationin \citep{Das:2011} show that a pile-up compression (PUC) of coronal plasma may form between coronal shocks and the CMEs behind them. This occurred in the simulation whenever the CME expanded faster laterally than radially. Their interpretation is that as a CME expands fast laterally, plasma piles in front of it in a `sheath' behind the shock \citep{Opher:2010}. Additionaly, there was no significant temperature increase in the PUC in the simulations, consistent with our DEM results.

Comparing the intensities of the June 13 wave with results from \citet[their Fig.~4]{Patsourakos:2010}, we find the wave began increasing brightness significantly towards 05:38~UT, roughly coincident with the maximum speed of the CME. Even as the CME bubble aspect ratio reduced to $~$1, the overall wave brightness increased, peaking around 05:42~UT (after that the wave begins to disappear from the AIA FOV). Since the waves are quite dim, it was difficult to obtain quantitative observations. Future work will elucidate the temporal connection between lateral overexpansion and PUC formation. However, the DEM result of no significant temperature change support the modeling findings of a plasma compression sheath behind the shock from \citet{Das:2011}.

\subsection{Importance of the Magnetic Geometry for Particle Acceleration and Release}

\begin{figure}
\epsscale{1.0}
\plotone{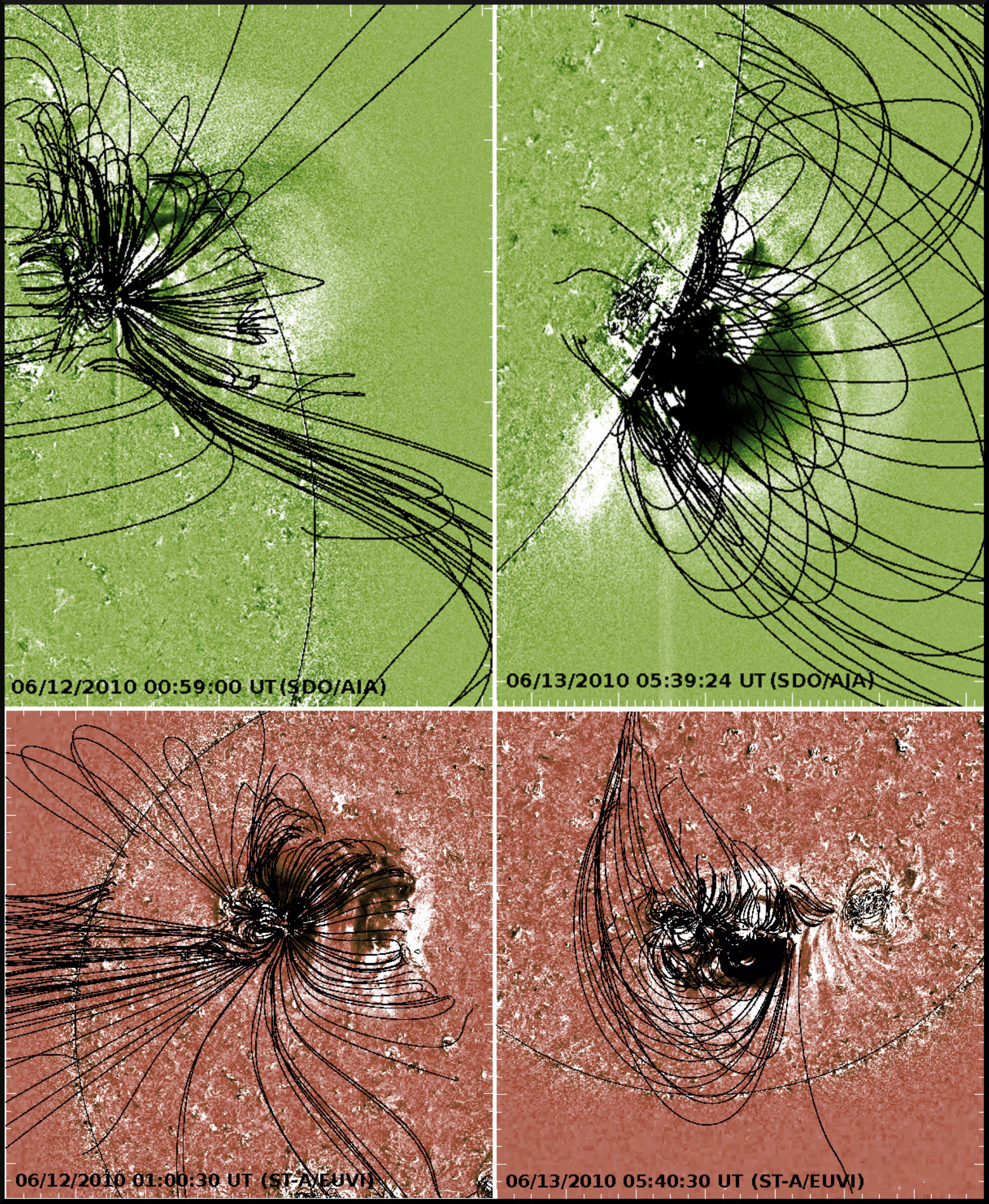}
\caption{SDO/AIA (top panels) and STEREO-Ahead/EUVI (bottom panels) base difference images during the June 12 (left) and 13 (right) events. The PFSS model coronal fields are overlaid to show the topology in which the waves/shocks propagated.\label{fig4}}
\end{figure}

Figure \ref{fig4} shows SDO/AIA (green) and STEREO-Ahead/EUVI \citep[red;][]{Howard:2008} difference images during both events, with a magnetic potential field source surface\citep[PFSS;][]{Schrijver:2003} model overlaid. On June 12 (left, top and bottom) the field geometry above the AR was very open, so particles were free to escape into interplanetary space as they gained energy. However, the complex magnetic topology does not allow us to address the possible sites of particle acceleration, and thus the discrepancy in positions and velocities between the EUV wave and radio shock.


On June 13 (Fig.~4, right panel), the magnetic geometry above the AR was much more closed - the shock might have been quasi-perpendicular at its nose, accelerating particles more effectively there (evidenced by the radio emission bands positioned in front of and behind the AIA wavefront). However, DEM analysis shows it was weak, so the 1 AU impulsive proton fluxes higher than 8~MeV did not increase above the already elevated levels (Fig.~\ref{fig1}, panel~E).

\section{SUMMARY}

We have presented observations of two western off-limb coronal waves in very high-cadence EUV imaging data. The waves were associated with metric type II bursts and significant increases in proton fluxes observed at 1 AU. We characterized the wave events in relation to the elevated particle fluxes at 1 AU. Our findings are:
1) The June 12 and 13, 2010 waves were large-scale, dome-like off-limb coronal transients, seen in EUV light. Enhanced emission sheaths followed the wavefronts.
2) The June 12 wave has a high initial speed ($\sim1287$~km~s$^{-1}$), but without a discernible driver, supported by a high average deceleration rate($\sim-1170$~m~s$^{-2}$).Similar behavior was observed in radio shock emission, although a discrepancy is clear between wave/shock positions and velocities. This might signify a more complex relation between the shock and wave, or alternatively, that the electron density model used for the radio data does not apply in this case. The June 13 wave started slower ($\sim736$~km~s$^{-1}$), but had a clear CME driver behind it sustaining its propagation, and consequently, a lower deceleration rate ($\sim-800$~m~s$^{-2}$).
3) DEM analysis of the June 13 wave event shows the enhanced emission was likely due to a density increase in the sheath behind the shock, and not to a temperature increase. We deduce from the emission measures ratio a density jump of at least $\sim1.12$.
4) EUV, radio, and in-situ observations, combined with a potential magnetic field model, reveal differences in the two events in terms of the possible field-to-shock orientation. In our interpretation, a more open field geometry of the June 12 event allowed protons accelerated impulsively (to $\sim50$~MeV) to escape quickly into interplanetary space. A closed field geometry during the June 13 event is supported by radio observations indicating the shock was effective in accelerating electrons at its nose, although proton fluxes above $\sim~8$~MeV at 1 AU did not increase appreciably.

The mechanisms of shock formation in the low corona are still under considerable debate. However, it seems that shocks do form low in the corona, and they are able to accelerate particles. The newly-introduced capability for multi-wavelength ultra-high cadence EUV observations of transients in the corona with SDO/AIA enables studying their dynamics in great detail. Based on our findings, the magnetic field geometry is important both for accelerating particles, and for their release into interplanetary space. Future work will involve analyzing multiple events and associated in-situ particle fluxes from multiple spacecraft, in order to constrain remote EUV wave observables significant for particle acceleration in the corona.


\acknowledgments
We acknowledge support under AIA subcontract SP02H1701R from Lockheed-Martin and NASA LWS EMMREM project NNX07AC14G. We thank David Long, Maher Dayeh, Marc De Rosa, Steve Saar, and Suli Ma for help and discussions.

\clearpage

\begin{deluxetable}{cccccc}

\tabletypesize{\scriptsize}
\tablecaption{EUV wave/radio shock kinematics, June 12 and 13, 2nd order fits\label{table1}}
\tablewidth{0pt}
\tablehead{
\colhead{Time} & \colhead{Channel/profile\tablenotemark{a} } & \colhead{Initial Velocity (km s$^{-1}$)} &
\colhead{Acceleration(km s$^{-2}$)} }

\startdata
06/12 00:56 & 193/II & 1169.34$\pm$29.31 & -0.88$\pm$0.16\\
06/12 00:56 & 193/III & 1381.59$\pm$32.70 & -1.29$\pm$0.17\\
06/12 00:56 & 211/II & 1180.25$\pm$31.67 & -0.95$\pm$0.17\\
06/12 00:56 & 211/III & 1418.08$\pm$31.56 & -1.55$\pm$0.17\\
\tableline
{\bf 06/12 00:56} & {\bf 193/AVG\tablenotemark{b}} & {\bf 1275.46$\pm$43.91} & {\bf -1.09$\pm$0.23}\\
{\bf 06/12 00:56} & {\bf 211/AVG} & {\bf 1299.16$\pm$44.71} & {\bf -1.25$\pm$0.24}\\
\tableline
\\
06/13 05:37 & 193/I & 774.29$\pm$18.90 & -0.97$\pm$0.18\\
06/13 05:37 & 193/II & 688.86$\pm$11.93 & -0.45$\pm$0.11\\
06/13 05:37 & 211/I & 791.89$\pm$25.30 & -1.15$\pm$0.24\\
06/13 05:37 & 211/II & 691.31$\pm$18.98 & -0.62$\pm$0.18\\
\tableline
{\bf 06/13 05:37} & {\bf 193/AVG} & {\bf 731.57$\pm$22.35} & {\bf -0.71$\pm$0.21}\\
{\bf 06/13 05:37} & {\bf 211/AVG} & {\bf 741.60$\pm$31.63} & {\bf -0.89$\pm$0.30}\\
\tableline
\\
\tableline
\tableline
06/12 00:57 & FUND & 2819.09$\pm$95.79 & -26.78$\pm$1.76\\
06/12 00:57 & HARM & 2905.60$\pm$147.92 & -46.79$\pm$5.30\\
06/13 05:39 & HARM & 589.54$\pm$27.21 & -0.53$\pm$0.23\\
06/13 05:38 & HARM & 610.93$\pm$14.97 & 0.28$\pm$0.11\\

\enddata
\tablecomments{Measurements for the 06/12 event started at 00:56 UT, for the 06/13 event - at 05:37 UT - the times we were able to first measure waves.}
\tablenotetext{a}{For radio measurements - emission type}
\tablenotetext{b}{Average of the profile measurements for that channel and event.}
\end{deluxetable}


\clearpage

\clearpage

\clearpage

\clearpage

\clearpage

\end{document}